\documentclass{iaus}
\usepackage{graphicx}

\title[Why SSS so weak in early-type galaxies?] 
{Why Supersoft X-ray Fluxes are So Weak\\ in Early-type Galaxies?\\
--A Clue to Type Ia SN Progenitors --}

\author[M.Kato]   
{Mariko Kato}

\affiliation{Keio University, \\ 4-1-1 Hiyoshi, Kouhoku-ku, 
Yokohama, 223-8521, Japan \\ email: {\tt mariko@educ.cc.keio.ac.jp} \\[\affilskip]
}

\pubyear{2011}
\volume{281}  
\jname{Binary Paths to the Explosions of Type Ia Supernovae}
\editors{R. Di Stefano \& M. Orio, eds.}
\begin{document}

\maketitle

\begin{abstract}
Supersoft X-ray (SSX) fluxes in early type galaxies provide an excellent test for
type Ia supernova (SN Ia) progenitors: Double degenerate 
(DD) scenario is believed to produce no SSXs except just before SN Ia explosion,
while single degenerate (SD) scenario produces in some phase of the symbiotic channel.
Recent observations of SSX flux of early type galaxies
show a remarkable agreement with theoretical prediction of SD scenario,
which is thus turn to be a strong support for SD scenario,
despite the original observation aimed at the opposite conclusion.
Here I explain why X-ray fluxes are so weak in early type of galaxies: 
(1) Candidate binaries in SD scenario become supersoft X-ray source
only in a short time on its way to an SN Ia explosion, 
because it spend a large part of lifetime as wind phase,  
(2) During the supersoft X-ray phase, symbiotic stars emit very weak SSX fluxes  
even if the WD is very massive.
It should be emphasized that 
SSX symbiotic stars are very rare and we need more observation to
understand their nature. 
\end{abstract}

\firstsection 
\section{Supersoft X-ray sources as a progenitor of SNe Ia in early type galaxies}

Type Ia supernovae (SNe Ia) play very important roles in astrophysics
as a standard candle to measure cosmological distances as well as
the production site of a large part of iron group elements.
However, the nature of SN Ia progenitors has not been clarified yet.
It has been commonly agreed that the exploding star
is a carbon-oxygen (C+O) white dwarf (WD) and the observed
features of SNe Ia are better explained by the Chandrasekhar mass
model than the sub-Chandrasekhar mass model.
However, there has been no clear observational indication
as to how the WD mass gets close enough to the Chandrasekhar mass for
carbon ignition ($M_{\rm Ia}= 1.38~M_\odot$), 
i.e., whether the WD accretes H/He-rich matter from
its binary companion [single degenerate (SD) scenario] or two C+O WDs
merge [double degenerate (DD) scenario].

\begin{table}
  \begin{center}
  \caption{Supersoft X-ray flux in early-type galaxies (0.3-0.7 keV).}
  \label{tab1}
 {\scriptsize
  \begin{tabular}{|l|l|c|c|c|c|c|c|}\hline 
{\bf Galaxy} &$N_{\rm WD,SSS}$ & $N_{\rm WD,SSS}$ &$l_{\rm X}^a$&$l_{\rm X}^a$  &$L_{\rm X,SSS}$ &$L_{\rm X,obs}$ &$L_{\rm X,SSS}$ \\ 
 &  &  & $10^{35}$ erg s$^{-1}$& $10^{35}$ erg s$^{-1}$ &$10^{37}$ erg s$^{-1}$  &$10^{37}$ erg s$^{-1}$&$10^{37}$ erg s$^{-1}$ \\
  & GB10$^b$ &HKN10$^b$&GB10 & HKN10 & GB10& GB10 & HKN10 \\
\hline
M32          &25 &3.7  &28& 3.1 &7.1& 0.15 & 0.12 \\
NGC3377      & 580 & 88  &47& 5.8 &270& 4.7 & 5.1 \\
M31 Bulge    & 1100 & 160 &21& 2.9 &230&6.3 & 4.7 \\
M105         & 1200 & 180 &46& 5.9 &550& 8.3 & 11 \\ 
NGC4278      &1600 & 240 &48& 7.2 &760& 15 & 17 \\
NGC3585      & 4400 & 660 &31& 3.5 &1400& 38 & 23 \\ \hline
  \end{tabular}
  }
 \end{center}
\vspace{1mm}
 \scriptsize{
 {\it Notes:}\\
  $^a$mean supersoft X-ray flux per source}\\
 $^b$ GB10 refers \cite[Gilfanov \& Bogd\'an (2010)]{GB10}, and 
HKN10 \cite[Hachisu et al. (2010)]{hac10kn}
\end{table}
The X-ray signature of these two possible paths are very different.
It is believed that no strong X-ray emission is expected from
the merger scenario until shortly before the SN Ia explosion.
On the other hand, the accreting WD becomes a supersoft X-ray source (SSS) long before
the SN Ia explosion. 
In order to constrain
progenitor models in early type galaxies,
\cite[Gilfanov \& Bogd\'an (2010)]{gil10}(hereafter, GB10) 
obtained supersoft X-ray luminosity $L_{\rm X, obs}$ for several 
early type galaxies as shown in the 7th column of Table \ref{tab1}. 
They concluded that these fluxes are much smaller than those expected from 
the SD scenario, thus, SD scenario is not the major path to SNe Ia. 
On the other hand, \cite[Hachisu et al. (2010)]{hac10kn} (hereafter, HKN10) 
argued, based on the same data of GB10, that the observed X-ray luminosities 
are very consistent with the SD scenario and thus, it is  
a strong observational support for SD scenario.
%
In this short report 
I will address what brings the difference between the two estimates by GB10 and 
HKN10 as well as current problems that must be studied in the future. 

\begin{figure}[b]
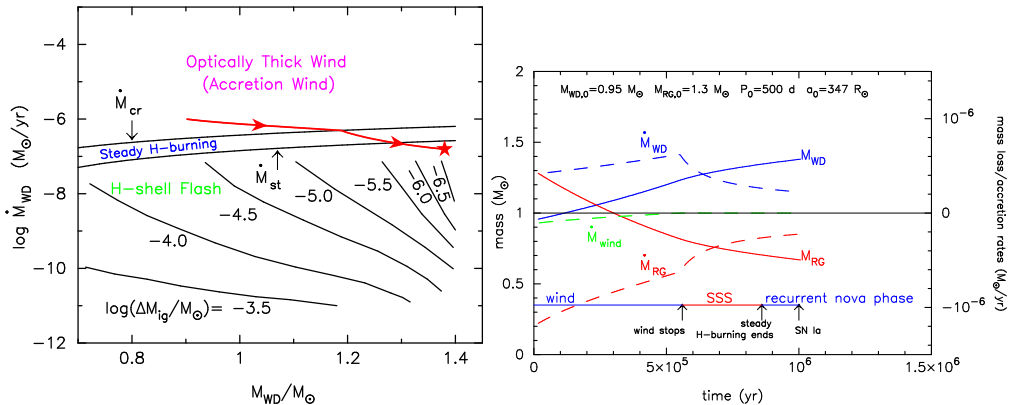

 \includegraphics[width=2.5in]{kato.sss.fig1a.epsi} 
 \includegraphics[width=2.7in]{kato.sss.fig1b.epsi}
 \caption{(left)
A typical evolutionary path ({\it red solid}) of a SN Ia progenitor
(in SD scenario) on the map of response of WDs to mass accretion rate.
Starting from the accretion wind phase, in which strong optically thick
winds blow from the WD, the binary enters the supersoft X-ray phase,
which is a narrow region between the accretion rate is $\dot M_{\rm cr} >
\dot M_{\rm WD} > \dot M_{\rm st}$.
When the mass accretion rate decreases, to less than the value to keep steady
hydrogen-burning ($\dot M_{\rm st}$), the binary enters the recurrent nova 
(very weak shell flash) phase. The WD explodes at the star mark as a SN Ia.
This figure is taken from \cite[Hachisu et al. (2010)]{hac10kn}.
(right) Evolution of a WD + RG system toward the Chandrasekhar mass limit. 
The mass transfer rate $\dot M_{\rm RG}$ from the RG
in an early phase exceeds $\dot M_{\rm cr}$. Thus, the WD blows optically thick winds
until $t = P_{\rm wind} \sim 5.5 \times 10^5$ yr and then
enters a SSS phase. As the mass accretion rate decreases, it
enters recurrent nova phase and reaches to the Chandrasekhar mass limit
at $ \sim 10^6$ yr.
}
 \label{fig1}
\end{figure}

\section{Supersoft X-ray phase in binary evolution of SD scenario}

The most important difference between GB10 and HKN10 is the number of WDs  in SSS phase. 
GB10 derived expected SN Ia rate for each galaxy,  
based on the K-band luminosity and assumed SN Ia rate per K-band luminosity.   
Then, GB10 derived the number of accreting WDs, $N_{\rm WD,SSS}$, 
assuming that all the WDs always stay in the SSS phase
before a SN Ia explosion 
(e.g, it takes $2 \times 10^6$ yrs for a 1.2 $M_\odot$ WD to increase its mass to the 
Chandrasekhar mass with accretion rate of 1 $\times 10^{-7} M_\odot$ yr$^{-1}$).  

This assumption, that a WD is always in the SSS phase, is however, very unlikely.  
In canonical SD scenario (e.g. \cite[Hachisu et al. 1999]{hkn99}), accreting WDs usually spend 
a large fraction of the lifetime in the
optically thick wind phase, then undergo the SSS phase, and finally enter the RN phase. 
Figure \ref{fig1} (left) shows a typical evolutional path of mass-accreting WD 
by the arrows, and Figure \ref{fig1} (right) demonstrates that SSS phase 
is relatively short in the total lifetime. 
In binary evolution calculations, the secondary star decreases its  
mass because of mass transfer, and then the mass transfer rate to the WD 
naturally decreases with time. 
Therefore, the WD evolves from the SSS phase to RN phase as the WD mass increases. 
GB10's assumed that WDs always evolve along the narrow strip of 'steady H-burning'  
(i.e., the mass accretion rate increases with time), but this is very artificial 
and highly unlikely.

HKN10 followed evolution of a number of binaries with different binary parameters 
and found that the SSS phase is 
as short as $P_{\rm SSS} \sim 2.5^{+0.9}_{-1.8} \times 10^5$ yr. Using this value HKN10 
recalculated $N_{\rm WD,SSS}$ with the same K-band luminosity and 
SN rate as GB10, which is shown in the third column of Table \ref{tab1}. 
Comparing with the GB10's estimates in the 2nd column, the number is reduced by factor of 8.

\section{Symbiotic stars are very dark supersoft X-ray sources}

Second discrepancy between GB10 and HKN10 is the 
``typical supersoft X-ray flux'' of candidate symbiotic stars. GB10 assumed absorbed 
X-ray flux $l_{\rm X}$ to be 2-5 $\times 10^{36}$ erg s$^{-1}$ per source as 
in the 4th column of Table \ref{tab1}. 
These values are consistent with theoretical estimates if (1) the WD luminosity is 
$L \sim 10^{38}$ erg~s$^{-1}$,  (2) the surface temperature is $T_{\rm eff}=45$ eV, 
(3) energy band of 0.3-0.7 keV,  and (4) no absorption except 
the Galactic one (i.e., no intrinsic absorption of symbiotic stars and 
no internal absorption within the early type galaxies).  
On the other hand, HKN10 adopted observation-based values, 
3-7 $\times 10^{35}$ erg s$^{-1}$ as listed in the 5th column of Table \ref{tab1};   
these small values are derived from arithmetic mean of absorbed fluxes of SMC3 and Lin358 
with corrected absorptions. 
 
Symbiotic stars are rare supersoft X-ray sources (e.g. \cite{mue97,sok11}). 
There are no strong supersoft X-ray symbiotic stars in the Galaxy and LMC except 
the one of metal poor star. All the LMC steady H-burning WDs are close binaries. 
In the SMC, two supersoft X-ray symbiotic stars are known: SMC3 and Lin358. 
SMC3 is the exceptionally bright SSS among the symbiotics in our Galaxy/the local group 
(\cite{mue97}), whereas Lin358 is a weak SSS. 
As SMC stars are metal poor, these observational facts suggest a possibility that 
symbiotics become bright SSS only in the metal-poor circumstance 
(J. Miko\l ajewska 2011: private 
communication at the conference). If this is the case, one cannot expect a larger number 
of bright SSS symbiotics in early type galaxies where metal is not poor. 
Even if not, it is very unlikely that all the galaxies in Table \ref{tab1} host 
a large number of symbiotic stars much brighter than the extraordinary bright X-ray 
symbiotic star, SMC3.

Table \ref{tab1} also shows the predicted supersoft X-ray fluxes for each galaxy, 
 $L_{\rm X, SSS}$, which are obtained from the X-ray flux per source ($l_{\rm X}$) 
multiplied by the number of sources ($N_{\rm WD,SSS}$). 
Estimates by GB10 (6th column of Table \ref{tab1}) are 40-70 times larger than 
those by HKN10 (the last column). 
We see the values by HKN10 are quite consistent with the observational value,  
$L_{\rm X, obs}$ in 7th column.

It should be stressed that we don't know much about symbiotic stars, in 
steady burning, massive WDs, except the fact of no stronger SSS symbiotics 
than SMC3 in nearby galaxies. 
Then a question arises; why symbiotic stars are so dark in supersoft X-rays? 
In other words, what was inappropriate among the conditions (1)-(4) listed above?
 
One possibility is (4) absorption. Symbiotic binaries could have 
high intrinsic absorption (\cite{sok11,stu11}).  
Another possibility is (2) temperature. For a low surface temperature,  
expected X-ray flux in a given energy band is sensitively reduced much 
due to low transparency. 
In symbiotic binaries, a WD accretes matter from the red giant companion 
possibly in wind-fed accretion rather than in disk-accretion. 
With a high mass-accretion rate, the WD atmosphere may be swollen up due to 
gravitational energy release and could have a much lower temperature than in 
steady burning WDs with disk-fed-accretion (like LMC SSSs). 
In fact the WD temperature of the two SMC symbiotics are in the lower side 
of temperature distribution of LMC SSSs (20-80 eV); 
$\sim 40$ eV (\cite{ori07}) and $\sim 33$ eV (\cite{stu11}) for SMC3,  
19.6 eV for Lin358 (\cite{kah06}).  If the envelope is further swollen up, 
the temperature becomes too low to emit X-rays. 
Therefore, the temperature of 45 eV (GB10's suggestion) seems to be 
an upper limit and a much higher temperature is unlikely in symbiotic stars.  
However, note that we have only two samples and need more 
statistics/observational information.

\section{'overproduction' of RNe/CNe in SD scenario?}

\cite{GB11} also claimed that observed number of novae in early type galaxies are 
much smaller than that expected from the SD scenario, and 
suggested that the SD scenario is not the major path to SNe Ia. 
My criticisms are as follows:
(1) They confused the recurrent novae (RNe) and classical novae (CNe), these two are in 
different binary evolutional paths. CNe are not the candidate of SN Ia, so 
the ratio of observed number of novae and expected number of RNe has no scientific meaning.  
Moreover, number of RNe are highly unknown, but many undetected RN candidates  
are suggested by \cite[Strope et al. (2010)]{str10}. 
(2) They overestimated the "expected number of RNe'' in the SD scenario
in the same logic as in Section 2, i.e., all the accreting WDs keep staying 
in the RN phase (not SSS phase, this time) during the lifetime.
(3) A large part of ``recurrent nova'' of Figure \ref{fig1} may not correspond to 
a typical RN like RS Oph, but may be like UV variables. 
Because shell flashes are so weak as well as a large duty cycle  
(e.g., \cite[Sion \& Starrfield 1986]{sio86}), these shell flashes may be very dark 
in optical wavelength and not identified as recurrent novae. 
Considering these uncertainties, one may reduce statistically expected number of "normal" RN 
(like RS Oph) by two order of magnitudes, and expected number of RN increase by 
a factor of 10, which makes the \cite{GB11}'s discrepancy almost vanish.


\end{document}